# Epsilon-Near-Zero Photonics Wires for Mid-Infrared Optical Lumped Circuitry

Runyu Liu, rliu31@illinois.edu, *Department of Electrical and Computer Engineering, Micro and Nanotechnology Lab, University of Illinois Urbana Champaign, 208 N. Wright St., Urbana, IL 61801, USA*

Christopher Roberts, Christopher_Roberts@student.uml.edu, *Department of Physics and Applied Physics, University of Massachusetts Lowell, One University Dr., Lowell, MA 01854, USA*

Yujun Zhong, yjzhong@illinois.edu, *Department of Electrical and Computer Engineering, Micro and Nanotechnology Lab, University of Illinois Urbana Champaign, 208 N. Wright St., Urbana, IL 61801, USA*

Viktor Podolskiy, viktor_podolskiy@uml.edu, *Department of Physics and Applied Physics, University of Massachusetts Lowell, One University Dr., Lowell, MA 01854, USA*

Daniel Wasserman (corresponding author), dwass@illinois.edu, *Department of Electrical and Computer Engineering, Micro and Nanotechnology Lab, University of Illinois Urbana Champaign, 208 N. Wright St., Urbana, IL 61801, USA*

*Phone: 217.333.9872*

*Fax: 217.244.6375*

**Abstract:** There has been recent interest in the development of optical analogues of lumped element circuitry, where optical elements act as effective optical inductors, capacitors, and resistors. Such optical circuitry requires the photonic equivalent of electrical wires, structures able carry optical frequency signals to and from the lumped circuit elements while simultaneously maintaining signal carrier wavelengths much larger than the size of the lumped elements. Here we demonstrate the design, fabrication, and characterization of hybrid metal/doped-semiconductor 'photonic wires' operating at optical frequencies with effective indices of propagation near-zero. Our samples are characterized by polarization and angle-dependent FTIR spectroscopy and modeled by finite element methods and rigorous coupled wave analysis. We demonstrate coupling to such photonic wires from free space, and show the effective wavelength of the excited mode to be approximately an order of magnitude larger than the free-space wavelength of our light. The operational length of the photonic wires approaches twice the free space wavelength, significantly longer than what is achievable with bulk epsilon near zero materials. The novel architecture utilized in our hybrid waveguides allows for significant design flexibility by control of the semiconductor material's optical properties and the sample geometry. In addition, by utilizing a semiconductor-based architecture, our photonic wires can be designed to monolithically integrate the optical equivalents of capacitive, inductive, and resistive lumped circuit elements, as well optoelectronic sources and detectors. As such, the demonstrated photonic wires have the potential to provide a key component, and a realistic framework, for the development of optical circuitry.

**Keywords:** Surface Waves, Metactronics, Plasmonics, Infrared

**Introduction.**

Metactronics, a new paradigm for engineering optical circuits and for optical information processing, has been recently proposed [1]. In this paradigm, light, not charge, plays the role of information carrier, with small (subwavelength) components playing the roles of lumped circuit elements. If a single material platform were available to realize all basic lumped circuit elements, the highly optimized apparatus for engineering electronic circuitry could be then applied to engineering optical circuits that would potentially operate at frequencies orders of magnitude above those of current electronic devices. Semiconductor technology has been at the forefront of developments in materials science, optics, and optoelectronics for the past several decades. As result, the optical properties of semiconductors can be engineered with relative ease, offering an opportunity to design and engineer the lumped optical elements required for the proposed paradigm of metactronics. Here we present a realization of perhaps the most essential metactronic component, the 'photonic wire'. Our design combines the doped semiconductor "designer metal" platform with a (near-perfect electrically conducting) metal-insulator slot waveguide operating near cut-off, to achieve photonic wires with both operational (propagation) lengths and local wavelengths longer than the free space wavelength of the optical signal wave. Most importantly, the demonstrated hybrid semiconductor/metal platform can leverage the versatility of epitaxial growth (of active devices and doping-controlled 'designer' materials) to realize both integrated optoelectronic devices and passive metactronic components on a single optical chip.

In electronics, the term "lumped elements" refers to those electrical components (resistors, capacitors, diodes, inductors, etc) that can be treated essentially as modular black boxes, and incorporated into circuit designs without requiring a microscopic model of the physics inside the black boxes. Typically such circuitry operates at electronic frequencies (radio or microwave) having wavelengths much larger than the size of the lumped elements. At these frequencies, photonic wires can be realized by operating microwave waveguides at cut-off [2], though the potential advantages of optical circuitry cannot be fully realized at these frequencies, already accessible with electronic components. Extending the concept of lumped elements to the optical frequency regime, however, is not trivial, and requires, among other things, the scaling of the size of the basic functional elements inside the optical circuits so that they can be treated as "lumped". Decades of advances in nano-fabrication and –characterization, as well as equivalent advances in simulations and modeling, have opened the door to the design and fabrication of nano-scale optical components. However, as feature sizes shrink to progressively subwavelength scales, control of the element geometry diminishes, while the expense of fabrication grows increasingly large.

The idea of lumped optical elements has been a topic of active research for several decades [3-5]. It was shown that from the point of view of bulk optics, a subwavelength particle with positive permittivity represents a capacitor; a particle with negative permittivity represents an inductor, while a particle with loss represents a resistor [1,3]. These optical lumped elements, in the proposed two-dimensional paradigm of metactronics [4], are linked by a "photonic wire", originally conceived as an air-filled guide with walls composed from hypothetical bulk material having vanishingly small permittivity, and thus having vanishingly small effective index [4,6].

Such materials are commonly referred to as epsilon near zero, or ENZ, materials, and have been predicted theoretically [7-9] as well as demonstrated experimentally, across a wide range of the electromagnetic (EM) spectrum [10-21]. Experimental realizations of ENZ materials have been demonstrated from visible to microwave frequencies, using composite materials designed to mimic bulk ENZ characteristics [10-16]. Alternatively, at optical frequencies, many plasmonic or phononic materials can be used as a homogeneous bulk ENZ platform [17-21]. However, the unique phenomena associated with such materials, as with the engineered materials which make up the larger fields of plasmonics and metamaterials, come at a price: optical loss. Thus, when used as the cladding of a metactronic optical wire, material absorption in realistic ENZ materials results in guided modes with $\tilde{n}_{eff} = n + i\kappa$ where $n \approx \kappa \geq 0.35$, limiting the operational "length" of the photonic wire. The mode profiles for such representative waveguides, fabricated in this case from highly doped semiconductor, operating at wavelengths where the permittivity of the doped semiconductor is nearest to zero ($\varepsilon_{InAs}(\lambda_{ENZ}) \approx 0$), and using realistic losses extracted from the literature, are shown in Figure 1, along with the variation in $n, \kappa$ for each waveguide as a function of geometry at ENZ wavelengths ($\lambda_{ENZ}$). Both the 1D slot and 2D core/cladding waveguides are shown, each having operational lengths (defined as the propagation length of the optical mode, $\delta = \lambda_o/2\pi\kappa$) less than a free-space wavelength and local wavelengths (defined as $\lambda = \lambda_o/n$) of approximately $\lambda = 2\lambda_o$. Even in the case of unrealistic (near lossless) materials, as soon as the 1D slot waveguide metactronic circuit (Fig. 1a) comes into contact with a dielectric, light would necessarily escape from the low-index optical wires into the higher-index vacuum. An attempt to address these challenges was made in [22] where a photonic wire based on a plasmonic waveguide was proposed and experimentally demonstrated. The effective index of this structure at $n \approx \kappa$, is $\tilde{n}_{eff} = 0.204 + 0.211i$. While an improvement over the ENZ-clad waveguides of Fig. 1, the structure still exhibits $\delta < \lambda_o$, and an effective wavelength $\lambda < 5\lambda_o$, due largely to the losses of traditional plasmonic (noble) metals at near-IR/Vis frequencies. Moreover, the material platform of this all-Au wire does not allow straightforward integration with other metactronic or optoelectronic components.

Here we propose and demonstrate a paradigm that takes advantage of the highly controllable optical response of designer ENZ materials made from doped semiconductors and which can thus potentially enable the design and fabrication of an integrated metactronic circuit. Our photonic wires are fabricated using a hybrid doped semiconductor-metal slot architecture, and can be spectrally tuned by controlling the material properties of the doped semiconductor component of the waveguide, as well as the waveguide geometry, increasing the length of the optical wire to multiple free-space wavelengths ($\delta \approx 2\lambda_o$). The demonstrated ENZ waveguides are of particular interest for the aforementioned nascent field of 'metactronics', serving as photonic wires able to carry signals with effective wavelengths much larger than the optical elements with which these modes interact ($\lambda > 10\lambda_o$). The use of semiconducting material as a plasmonic component allows for structures fabricated with epitaxial $\varepsilon > 0$ and $\varepsilon < 0$ materials to be grown directly into the waveguides, as well as the potential integration with active optoelectronic structures such as detectors and sources. Our waveguides are designed and modeled using both rigorous wave couple analysis (RCWA) and finite element methods, and characterized using angle- and polarization-dependent Fourier transform infrared (FTIR) spectroscopy. We demonstrate coupling from free space into our waveguiding photonic wire structures, analyze the excited modes in these waveguides, and discuss the challenges as well as the opportunities of the demonstrated photonic wire architecture.

**Materials and Methods.**

*Photonic Wire Design.* The ideal photonic wire should satisfy several important constraints. First, it should be isolated from the surroundings so as to minimize out-of-guide radiation leakage and parasitic coupling of signals to the surrounding optical environment. Second, the wire should be fabricated on a platform that would allow, in principle, fabrication of the other elements of photonic circuitry. Third, the structure (as well as the other optical elements) should be realizable with current fabrication protocols. Finally, the effective index (both real and imaginary) of the wire should be as low as possible. The latter condition results from the requirement that the ideal photonic wire not only demonstrates a dramatic extension of the local wavelength ($\lambda = \lambda_o/n \gg \lambda_o$) but also is able to propagate for distances longer than the free space wavelength ($\delta = \lambda_o/2\pi\kappa \gg \lambda_o$). Minimizing both $n$ and $\kappa$ provides the primary (significant) challenge to designing and demonstrating photonic wires for optical circuitry applications.

The constraints outlined above virtually eliminate the possibility of all-dielectric (ENZ-cladding) photonic wires, such as the designs shown in Fig. 1, due to the sub-wavelength operational length ($\delta < \lambda_o$) and because the radiation to be guided will leak out of such structures into the surrounding environment, which has $n \simeq 1$. However, were bound surface modes to be utilized, such leakage into the continuum could be prevented, or at least minimized. Furthermore, such modes would offer the opportunity to enhance the interaction between bound optical waves and any optical lumped circuitry integrated onto the waveguide base. While traditional polaritonic surface modes, such as the surface plasmon polaritons (SPPs) supported on semi-infinite planar metal/dielectric interfaces, do not offer dispersion relations with $n_{eff.} \approx 0$ regimes, the addition of lateral confinement to the SPP mode results in a waveguide with a frequency cut-off where $n_{eff.} \approx 0$. Below, we investigate several designs of photonic wires based on surface modes and compares their performance as far as the extension of the quasistatic limit is concerned. As a figure of merit, we use the value of the real part of the effective index at the wavelength where $Im\{\tilde{n}\} = Re\{\tilde{n}\}$, with smaller values of $Im\{\tilde{n}\} = Re\{\tilde{n}\}$ corresponding to modes with lower losses, larger effective wavelength, and larger operational length. The optical properties of the modes in our waveguide designs are modeled using finite element method (FEM) using eigenvalue/eigenmode analysis offered by the commercially available COMSOL Multiphysics package [23,23].

Fig. 2 presents the characteristics of four waveguide structures with the potential for serving as photonic wires at mid-IR frequencies. The first two structures shown are waveguides fabricated entirely from Au; an open (Fig. 2c) and capped (Fig. 2d) Au slot waveguide with an Air gap, similar to the structure of [22,22]. The latter two waveguides represent the 'hybrid' slot waveguides proposed and developed here, either open (Fig. 2e) or Au-capped (Fig. 2f) with a doped InAs base and Au sidewalls. At mid-IR wavelengths, the optical response of Au closely resembles that of a perfect electrical conductor (PEC), while the doped InAs represents a plasmonic material operating close to its plasma frequency [[24]24]. The PEC nature of the Au in the mid-IR accounts for the improved performance of the scaled all-Au waveguide in Fig. 2d when compared to its near-IR-VIS counterpart from Ref. 22. While the all-Au structures (Fig.2 c,d) behave similarly to PEC waveguides, operating at cut-off, the hybrid waveguides presented here (Figs. 2e,f), take advantage of a SPP mode supported by the air-doped semiconductor interface that is brought to cut-off by two PEC vertical sidewalls. Fig. 2a shows the dependence of $n(\lambda), \kappa(\lambda)$ for each of the four waveguide structures, demonstrating that values of $n \approx \kappa \leq 0.15$ are possible for all four designs presented, a marked improvement over the all-dielectric waveguide structures shown in Fig. 1 and the near-IR/VIS Au-waveguide of Ref. 22.

Despite similar performance in the $n \approx \kappa$ regime, the mode profiles of the all-Au and hybrid waveguides differ significantly. Fig. 2b shows the dependence of the $Im\{\tilde{n}\} = Re\{\tilde{n}\}$ value for each of the four waveguides as a function of $h$, the height of the waveguide sidewalls. The all-Au open waveguide (Fig. 2c) shows poor performance for small $h$, a result of light leakage from the from the poorly bound Au/Air surface mode, as can be seen in the mode profile shown in Fig. 2c. For large values of $h$, however, this waveguide gives low $Im\{\tilde{n}\} = Re\{\tilde{n}\}$ values, when the mode is more tightly bound by the ~PEC sidewalls. However, the values of $h > \lambda_o$ required to achieve low losses are not only difficult to fabricate, but approach the cross-sectional length scales of quasi-plane waves, preventing strong interaction with subwavelength optical lumped elements. Not surprisingly, the open hybrid waveguide modeled shows the often encountered plasmonic trade-off between mode confinement and propagation length. The hybrid waveguide shows slightly lower losses at small $h$, due to the stronger confinement arising from the plasmonic nature of the doped semiconductor at long wavelengths. At larger $h$, the hybrid waveguide shows higher losses than its all-Au counterpart, as confinement is provided by the plasmonic semiconductor, which results in a minimum $n \approx \kappa$ value that is largely independent of sidewall height. For realistic values of $h$ ($h \leq \lambda_o/2$), the uncapped hybrid waveguide slightly outperforms its all-Au counterpart, due the stronger confinement of the surface mode on the plasmonic semiconductor surface, and therefore the reduced leakage into the continuum.

The addition of an Au-cap to the modeled waveguides significantly improves the performance of both the all-Au and hybrid designs. Figs. 2e, f show the prospects for designing completely encapsulated photonic wires. It is seen that these structures have even longer operational length, when compared to the open slot waveguides. Using such a design, the hybrid waveguide demonstrates $Im\{\tilde{n}\} = Re\{\tilde{n}\} < 0.08$ with $h \leq \lambda_o/2$, giving $\delta \approx 2\lambda_o$, wire lengths potentially compatible with optical lumped element circuit designs. The hybrid capped waveguide slightly underperforms its all-Au capped counterpart (for which leakage has been eliminated due to the Au cap), though both structures show significant improvement in performance when compared to the ENZ-clad, near-IR/VIS all-Au, and uncapped slot waveguides. At the same time, the hybrid waveguides offer multiple additional design benefits. First, the dispersion of the hybrid mode can be controlled both by geometry (varying the width of the slot waveguide) and by material (control over the permittivity of the doped semiconductor during epitaxial growth) [25-27]. Second, because our waveguide is fabricated on an epitaxially-grown semiconductor base, both bulk semiconductor (doped and undoped), as well as optoelectronic active regions, can be grown epitaxially in the waveguide region, offering the opportunity for future integration of passive optical lumped elements and/or optoelectronic emitters and detectors with the optical wires.

In the following sections, we describe the fabrication of the photonic wires modeled in Fig. 2c, as well as the experimental set-up, and numerical simulations, developed to characterize the fabricated photonic wires.

*Photonic Wire Fabrication.* Our waveguide structures were fabricated on epitaxially grown highly-doped InAs layers, grown in an SVT Associates molecular beam epitaxy (MBE) system using a growth process described in depth in previous work [25,26]. Figure 3 shows the growth structure of the epitaxial material, as well as the normal incidence reflection spectrum from the epitaxial surface. From this spectrum, using a transfer matrix method (TMM) approach [28,29], the plasma frequency and scattering rate of the doped InAs can be extracted, allowing for the determination of the complex permittivity of the InAs, using the Drude model

$$\varepsilon(\omega) = \varepsilon_\infty \left(1 - \frac{\omega_p^2}{\omega^2 + i\omega\gamma}\right) \, , \, \omega_p^2 = \frac{ne^2}{\varepsilon_\infty \varepsilon_o m^*} \tag{1}$$

where $\varepsilon_\infty$ is the semiconductor's high frequency dielectric permittivity, $\gamma$ the free carrier scattering rate, and $\omega_p$ the doped semiconductor plasma frequency, which depends on the free carrier effective mass $m^*$ and the free carrier concentration $n$. The extracted complex permittivity of our epitaxially-grown, doped InAs, using the Drude model with a TMM fitting approach, is shown in Fig. 3b.

The waveguide fabrication process, for a single waveguide, is shown in Fig. 4a. Upon the epitaxial surface, we first pattern an Au grating with stripe width 12 μm, period 18 μm and thickness 50 nm using UV photolithography, followed by a metal deposition and lift-off process. Subsequently, a layer of SU-8 with thickness of ~6 μm is spin-coated over the surface of the sample. A second lithography step is used to pattern SU-8 stripes directly above the existing Au stripes. Next, a double angled metal evaporation (75nm Au at 31° from normal) is performed to coat the SU-8 stripe sidewalls, using the shadow effect to prevent metal deposition on the exposed semiconductor between the SU-8 stripes. A scanning electron micrograph (SEM) of a waveguide cross-section, after the metal deposition step, is shown in Fig. 4b. The sample was then placed face down on an Au-coated Silicon wafer, and the edge of the sample sealed with crystal bond wax. The GaAs substrate was removed using a selective wet etch (solution $NH_4OH : H_2O_2: H_2O = 8:24:128$), and the thin GaSb buffer removed using a different selective wet etch (solution $HCL : H_2O_2: H_2O = 100 : 1 : 100$), leaving the substrate side of the doped InAs layer exposed.

The result of the waveguide fabrication process is an array of hybrid slot waveguides with metal sidewalls and a doped InAs base. The metal sidewalls of the waveguides ensure that each waveguide is optically isolated from its neighboring waveguide structures, while the spatial density of the waveguides results in a stronger interaction of the waveguide modes with the large beam diameter of the mid-IR probe beam. Fig. 4a (viii) shows a schematic of the final fabricated structure directly bonded to a Au-coated carrier wafer. It should be noted that due to bowing of the epi- and carrier wafers, it is possible that a slight air-gap spacer exists between the two sample surfaces. The magnitude of this spacer gap may vary across the sample surface, an effect which would result in a broadening of our experimental data, as discussed below.

*Experimental Set-up and Numerical Calculations.* Our structure was characterized by angle and polarization dependent FTIR spectroscopy using the experimental set-up shown in Fig. 4c. Light from the internal source of the FTIR passes through two spatial filters designed to collimate and reduce the beam size of the light incident on the sample. A wire grid polarizer after the second spatial filter controls the polarization of the incident light. The sample is placed on a rotating stage to control the angle of the incident beam, and reflected light is collected by a parabolic mirror and focused onto a HgCdTe (MCT) detector mounted on a rotational stage concentric with the sample stage. Such a configuration is essentially the Kretchmann configuration used to measure coupling to plasmonic modes on thin metal films, except in this case we are coupling through an optically thin doped semiconductor layer (instead of a metal) to low-index hybrid waveguide modes (instead of higher index SPP modes). It is thus the TM polarized incident light, as shown in the inset of Fig. 4c, which is able to couple to the waveguide modes of interest. By varying incidence angle, we can control the wavevector of the incident light in the direction of waveguide propagation. Because dips in reflection correspond to coupling to these waveguide modes, tracking the spectral position of the dips as a function of incidence angle (mode wavevector) allows us to map the dispersion of the photonic wire modes. The magnitude of the reflection features will depend on coupling strength to the waveguides: thick n+ InAs layers will better confine the excited modes, but weaken coupling, while thin InAs layers will improve coupling, but to weakly bound modes. The thickness of the n+ InAs was chosen to be 550nm, thick enough to confine the waveguide modes but thin enough to allow coupling. As incidence angle decreases (towards near-normal incident light) coupling to the true ENZ-like modes is enabled. However, the component of the electric field overlapping the waveguide mode's field profile decreases, weakening coupling. Thus, our experimental reflection features would be expected to be strongest for larger angles of incidence.

Our waveguide structures are modeled using custom-built rigorous wave coupled analysis (RCWA) software [30]. The RCWA technique originally proposed in [31], can model periodic structures imposing a Bloch-wave-periodicity to the system. This technique allows a rigorous calculation of the reflection of the system. TE- and TM-polarized reflection from our waveguide was simulated for the experimental configuration shown in Fig. 2c, for the waveguide geometry shown in Fig. 4a(viii), having a waveguide width $w = 6.15 \mu m$, sidewall height $h = 6 \mu m$, and periodicity $\Lambda = 18 \mu m$. We use the permittivity of Au from the literature [32] and the permittivity of the doped InAs extracted from the reflection measurements of epitaxially-grown material, shown in Fig. 3. Finally, we assume a spacing layer of $1.5 \mu m$ between the top of the waveguide sidewalls and the Au-coated carrier wafer, which we discuss further in the subsequent section.

**Results and Discussion.**

The RCWA simulations of TE- and TM-polarized reflection from our fabricated waveguide structures, as a function of incidence angle, are shown in Figs. 5 (a) and (b), respectively. Here we see distinct dips in the reflection spectra corresponding to coupling to modes in our fabricated waveguide arrays. The dashed line in Fig. 5b shows the dispersion of the hybrid waveguide mode of interest, shown in Fig.2f, calculated with the finite element method, and superimposed on Fig. 5b using the momentum matching equation

$$\frac{2\pi}{\lambda_o} Re(\tilde{n}) = k_z = \frac{2\pi}{\lambda_o} \sin \theta_i \tag{2}$$

with $k_z$ being the wavevector of the mode in the direction of the mode propagation, $\lambda_0$ being the free-space wavelength, and $\theta_i$ being angle of incidence. As expected, this mode is only present in TM-polarized excitation, and exhibits ENZ behavior for $12 < \lambda_0 < 13 \mu m$. Coupling to the ENZ mode weakens for decreasing angles of incidence, resulting from the aforementioned weaker overlap of the incident TM-polarized E-field with the excited mode. The numerical solutions of Maxwell equations reveal the presence of several other modes in the structure at $\lambda_0 \approx 12 \mu m$. Our calculations show that these modes are either (i) volumetric modes that do not have significant overlap with InAs substrate and thus cannot be modulated by doping or (ii) surface-guided modes that are supported by the bottom (unconstrained) InAs-air interface that would not exist in realistic photonic circuits.

Figures 6 (a) and (b) show the experimental results from the fabricated samples for TE- and TM-polarized incident light. The experimental results largely match the simulated reflection spectra of Fig. 5, with coupling to the ENZ-like mode observed across a similar range of wavelengths and incidence angles.

Additional coupling features are seen in the wavelength range between $\lambda_o = 8 - 10 \mu m$ for incidence angles $\theta < 40°$ in both TE and TM polarized data, for both our experimental and simulation results. Similar to the extra modes at $\lambda_0 \approx 12 \ \mu m$, these modes are identified as volumetric waveguide modes, which also will approach effective indices $n_{eff} \approx 0$ near cut-off. However, these modes are not tied to the semiconductor surface (being centered at the waveguide core), and therefore will not strongly interact with any structures fabricated upon the waveguide base. The inset in Fig. 5a shows the mode profile of such a volumetric mode, demonstrating the weak overlap of the mode with the surface of the InAs. Notably, these modes only appear in the capped waveguide simulations; without the Au cap, the volumetric modes are not bound. The appearance of these modes in the experimental data indicates that the Au-coated surface of the carrier wafer affects the optical properties of the fabricated waveguides. The Au-cap has the strongest effect on the volumetric modes which otherwise would not be bound to the waveguide core. However, as demonstrated in Fig. 2, the Au-cap will also significantly improve the performance of the ENZ modes (without drastically altering the mode profiles). The presence of the volumetric modes in our experimental data also allows for an estimate of the average spacer between the top of our waveguide sidewalls and the Au-coated carrier wafer, which we set at $1.5 \mu m$ for our simulations.

While our simulations and experimental results show good agreement, the experimental data of Fig. 6 can be seen to have a lower baseline reflectivity, as well as slightly broadened coupling features when compared to numerical calculations. The former effect is most likely due to fabrication-related imperfections, including surface roughness and/or oxidation of the exposed, wet-etched n-doped InAs surface upon which our probe beam is incident. We attribute the broadening of the spectral features observed in our experimental results primarily to the variation in the waveguide geometry across length scales commensurate with the probe beam diameter. Small changes in sidewall height, as well as the spacing between the top of the sidewall and the Au-coated carrier wafer can result in slight shifts of the observed spectral features. The wide probe beam diameter (required to maintain minimal angular divergence of the incident beam), ensures that our experiment samples a large number of waveguides, broadening the observed spectral features.

Figure 7 shows a comparison of the propagation lengths and cross-sectional mode profiles (in the direction of propagation) for the photonic wires discussed in this work and the ENZ cladding/air core waveguides proposed in [11], all modeled using realistic losses given in the literature [32] and/or extracted from our optical characterization of the doped semiconductor (Fig. 3). It is seen that the ENZ-clad guides are highly lossy, and decay over a length scale much shorter than a wavelength. It is also seen that while these waveguides can be designed to have small crossections, the electric field of the bound mode extends over macroscopic distance outside the waveguide core.

The all-Au slot waveguides, both open and capped, offer propagation lengths larger than a free-space wavelength. However, these structures support modes which are either weakly bound to the waveguide base (open waveguide) or alternatively, split between the top and bottom surface of the waveguide (capped waveguide). Moreover, the all-Au architecture does not allow for direct integration with optoelectronic devices or epitaxially-grown structures on the waveguide base. The hybrid photonic wires, the subject of this work, not only demonstrate potential wire lengths over 2 free space wavelengths ($\delta > 2\lambda_o$), but also local wavelengths an order of magnitude larger than the signal's free space wavelength ($\lambda > 10\lambda_o$), while at the same time allowing for direct integration of epitaxially-grown materials and devices at the base of the waveguide, where the mode profile is strongest. The lateral confinement of the field in the hybrid and Au-based guides is comparable or better than that of the guides with ENZ cladding.

The waveguides fabricated and characterized in this work utilize an optically thin ($d \ll \lambda_o$) waveguide base in order to allow optical coupling to, and thus characterization of, the photonic wires using free space radiation. Even so, optical coupling to the most ENZ-like modes of our wires (near-normal incidence) is weak, resulting from the aforementioned weak overlap of the incident light and waveguide modes electric fields. Though coupling is weak to the most ENZ-like modes, those modes are clearly supported in our waveguides (as evidenced by the excellent agreement between our simulations and experimental results). A photonic wire integrated into a metatronic circuit would presumably use a thicker doped InAs base (or alternatively cap the base with a layer of Au) in order to prevent light leakage from the photonic wire, and could be excited by a source embedded in the wire itself.

The demonstrated photonic wires operate in the mid-IR wavelength range at frequencies well above those of electronic of lumped element circuits, key to realizing the potential benefits associated with the paradigm of metatronics. However, the hybrid design utilized here does offer prospects for even higher frequency operation (potentially in the near-IR wavelength range) utilizing transparent conducting oxides or nitrides [refs[33]] as the plasmonic layers, combined again with similar Au sidewalls. Such structures could offer operation at telecom frequencies, where a well-established optical infrastructure exists. However, such a design would prevent monolithic integration of mectatronic components with a single epitaxial growth, and could suffer from additional material losses (both in the conducting oxides and nitrides, and the Au layers), making the extension of the demonstrated photonic wires to even higher frequencies an exciting challenge for future work.

**Conclusions.**

In summary, we have demonstrated epsilon-near-zero photonic wires, operating at mid-IR frequencies, with potential for metatronic applications. Our photonic wires are realized using surface plasmon waveguides brought to cut-off by a metal-insulator-metal slot waveguide structure. The photonic wires are fabricated on epitaxially-grown highly doped semiconducting layers, which serve as the plasmonic material base for the hybrid waveguide geometry, and are characterized experimentally using angle- and polarization-dependent infrared reflection spectroscopy. Good agreement is shown between our experimental results and RCWA calculations of the fabricated waveguides. The performance of our photonic wires is compared to the previously proposed 'all-dielectric' waveguide structures utilizing bulk ENZ cladding with realistic losses, as well as to all-Au slot waveguides (both at near-IR and mid-IR wavelengths). Our photonic wires improve on the ENZ-clad waveguide structures, demonstrating a significant enhancement in propagation length, as well as a dramatic improvement in the local wavelength of the guided mode. When compared to all-Au structures in the mid-IR, our photonic wires show comparable wavelength extension and propagation lengths, but offer the added advantage for potential integration with epitaxially-grown optical and optoelectronic structures and devices. With effective wavelengths of $10\lambda_o \approx 120\mu m$, our photonic wires offer the potential for straightforward integration of optical lumped elements having characteristic length scales $\ll \lambda_o$, and the experimental realization of metactronic circuitry at optical frequencies.

**Acknowledgments.**

The authors gratefully acknowledge funding support from the National Science Foundation (NSF) (1210398, 1209761), and the University of Illinois Post-Doctoral Drive Fellowship.

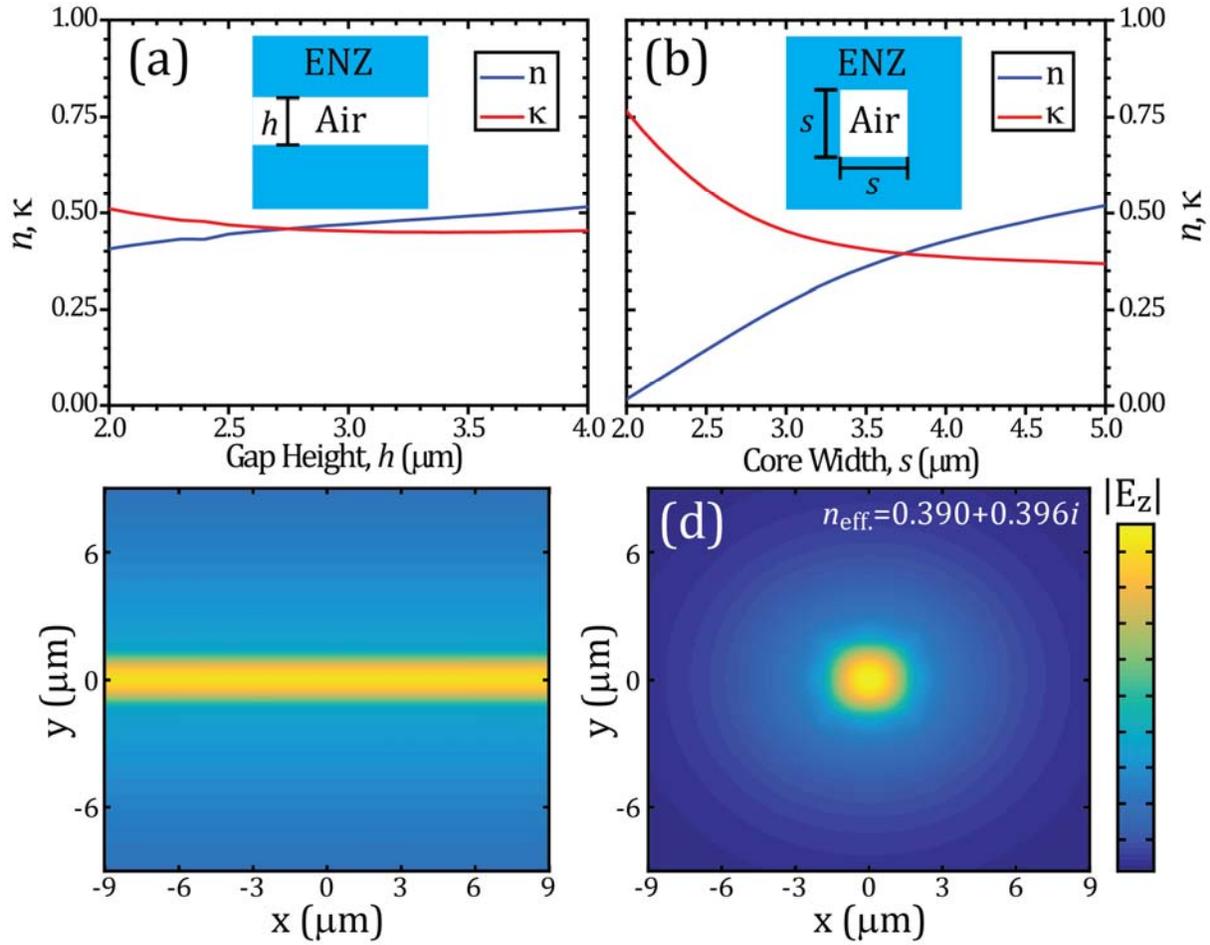

**Figure 1 | Simulations of ENZ-clad Photonics Wires.** Effective $n, \kappa$ for (a) 1D and (b) 2D ENZ cladding, air core waveguides as a function of (a) gap height $h$ and (b) core width $s$, modeled using realistic losses obtained from MBE grown n-doped InAs with $\varepsilon_{InAs}(\lambda = 6.9\mu m) = 0.0165 + 0.45i$. Insets show the basic structure used in our model. Cross-sectional field profiles ($|E_z|$) for (c) 1D and (d) 2D waveguides with (c) $h = 2.7\mu m$ and (d) $s = 3.7\mu m$, at the wavelength where $Im\{\tilde{n}\} = Re\{\tilde{n}\}$.

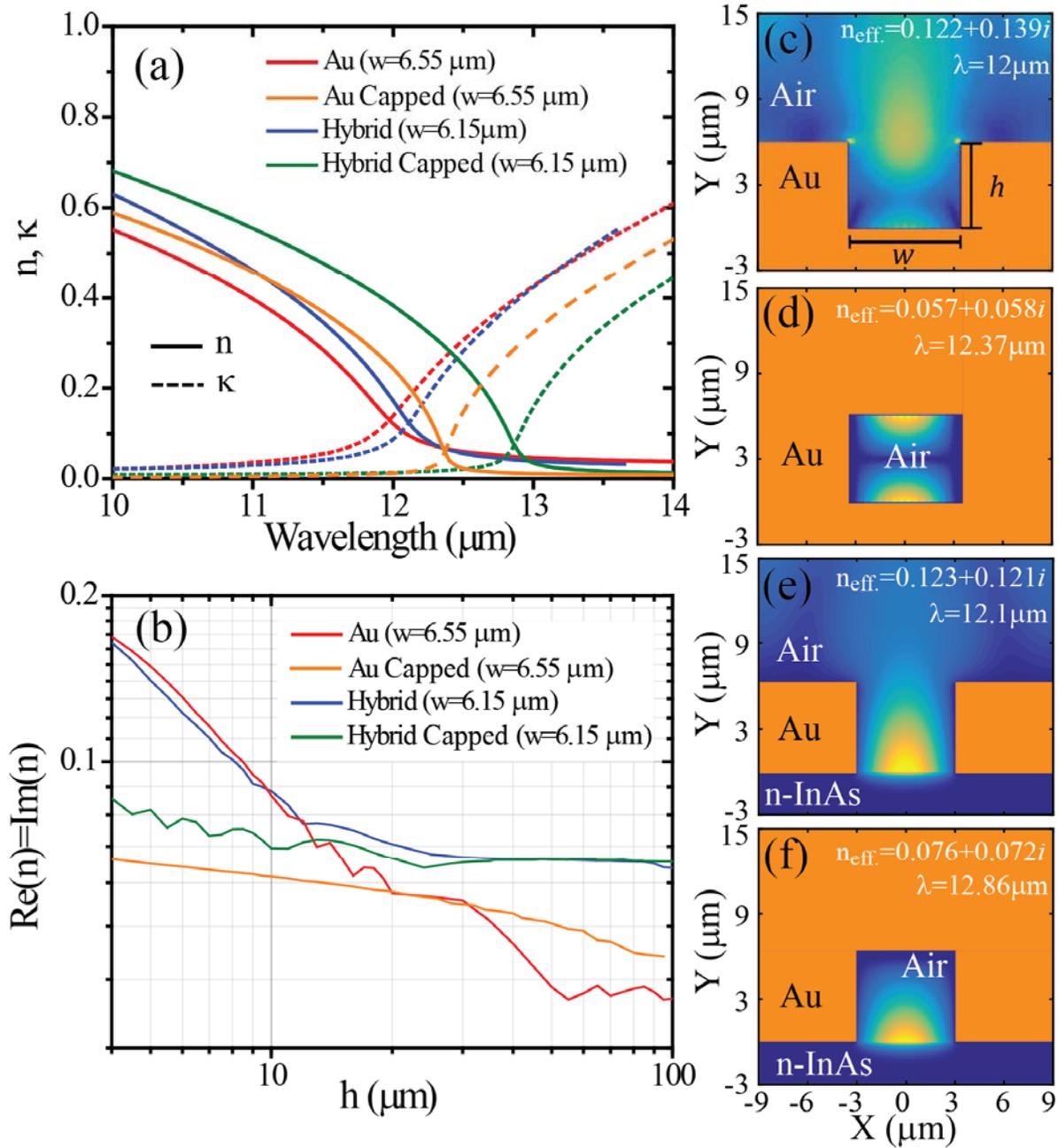

**Figure 2 | Simulations of All-Au and Hybrid Au/Plasmonic Photonic Wires.** (a) Modeled $n, \kappa$ vs. wavelength ($h = 6\mu m$) and (b) value of $Im\{\tilde{n}\} = Re\{\tilde{n}\}$ as a function of sidewall height $h$, for (c) all-Au (red), (d) all Au with Au cap (orange), (e) hybrid InAs/Au (blue) and (f) hybrid Au/InAs with Au cap (green) waveguides. Normalized mode profiles ($|E_z|$) at $Im\{\tilde{n}\} = Re\{\tilde{n}\}$ are shown in (c-f).

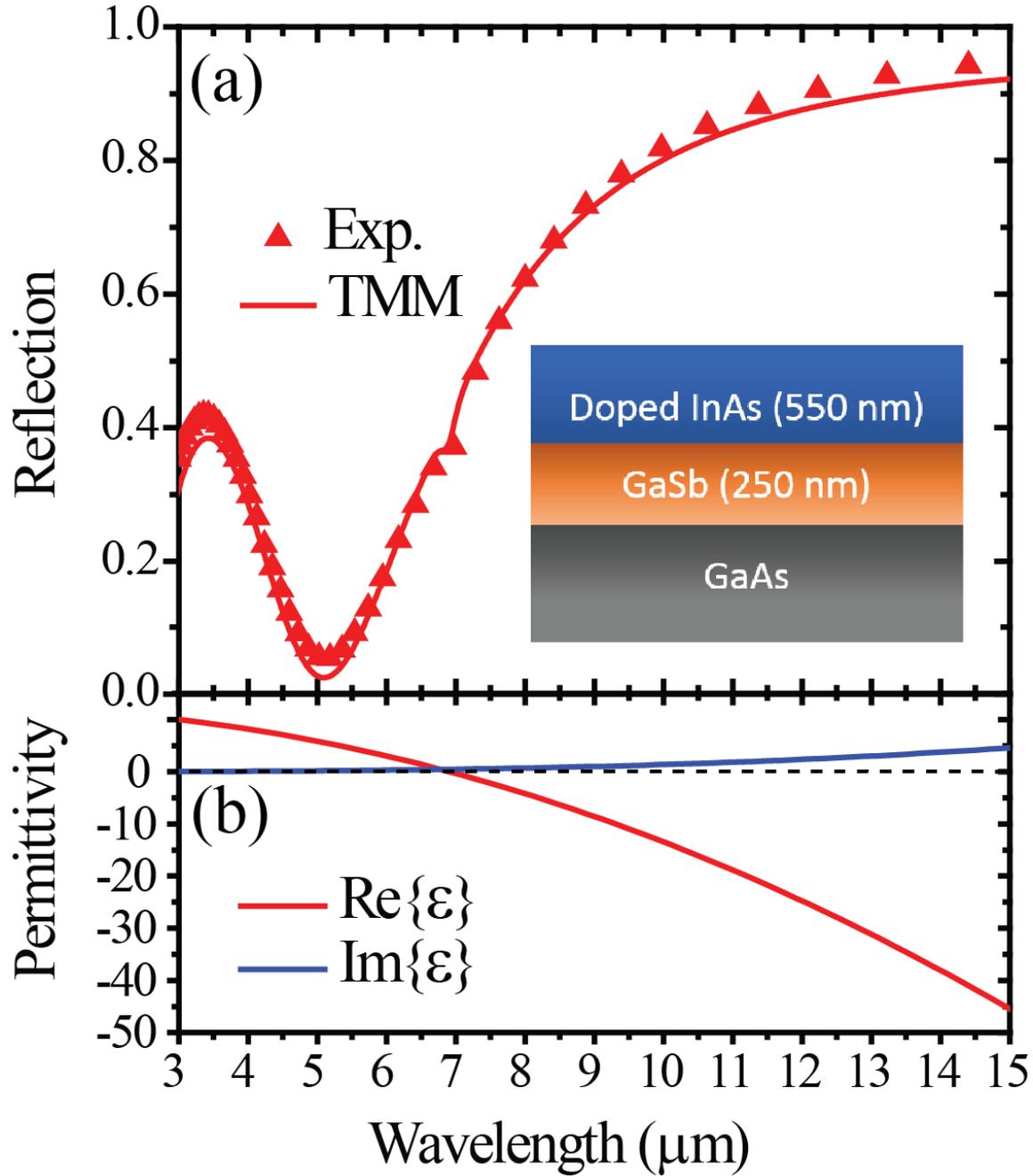

**Figure 3 | Characterization of Epitaxially-Grown Plasmonic Material.** (a) TMM-modeled (solid red) and experimental (red triangle) reflection spectra for as-grown epitaxial structure. Inset shows epitaxial layer structure. (b) Extracted complex permittivity of the doped InAs layer. Fitting parameters: $\varepsilon_{\infty,GaAs} = 10.89$, $\varepsilon_{\infty,GaSb} = 14.45$, $\varepsilon_{\infty,InAs} = 12.3$, $\lambda_{p,InAs} = 6.9 \mu m$, and $\gamma_{InAs} = 1 \times 10^{13} s^{-1}$.

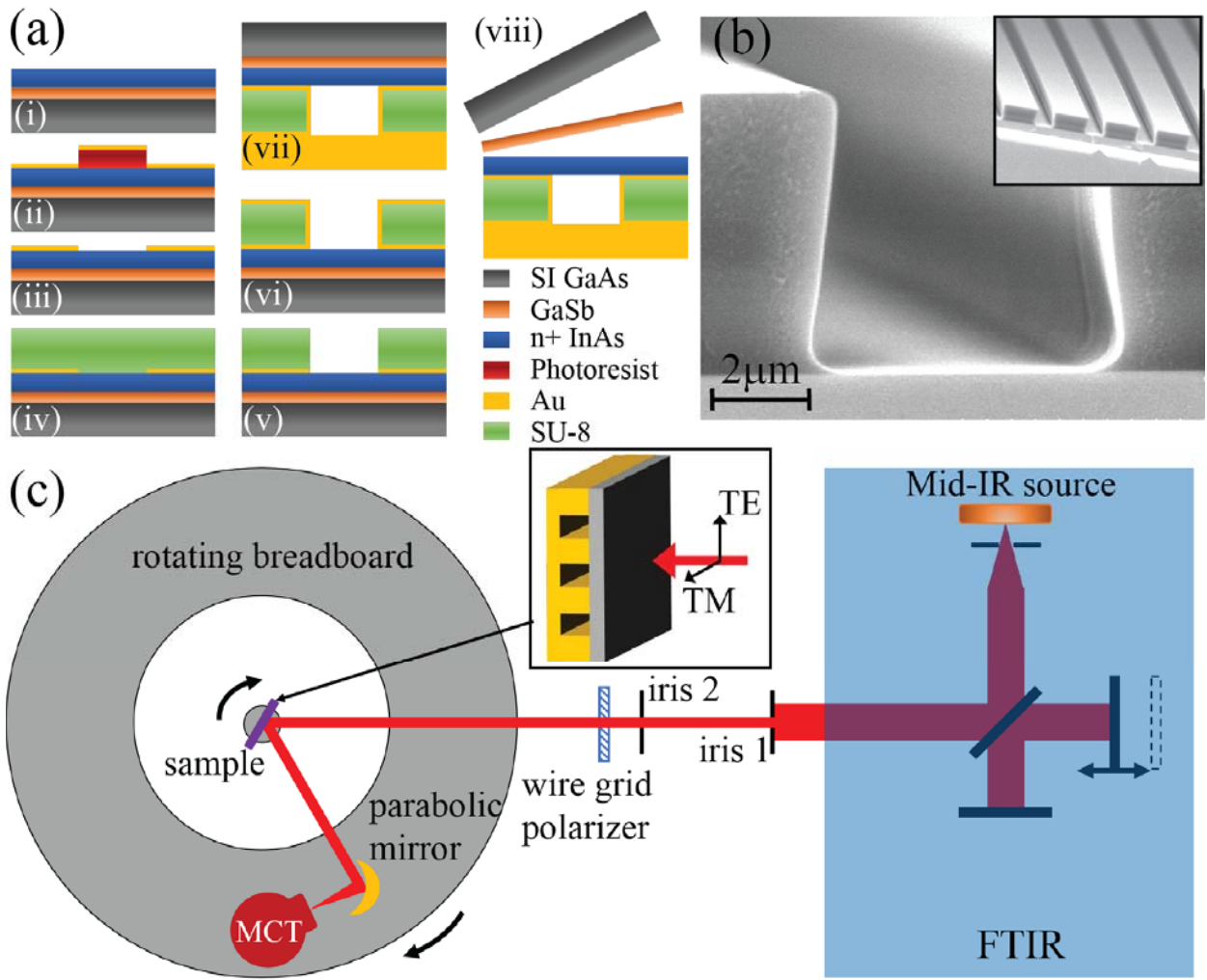

**Figure 4 | Fabrication Process Flow and Experimental Characterization Set-Up for Photonics Wires**. (a) Process flow for ENZ waveguide fabrication and (b) SEM image of the cross section of the ENZ waveguide, with inset showing expanded view of ENZ waveguide array. (c) Experimental setup for characterizing the angle- and polarization-dependent reflection from the ENZ waveguide array, with inset showing the sample orientation under test.

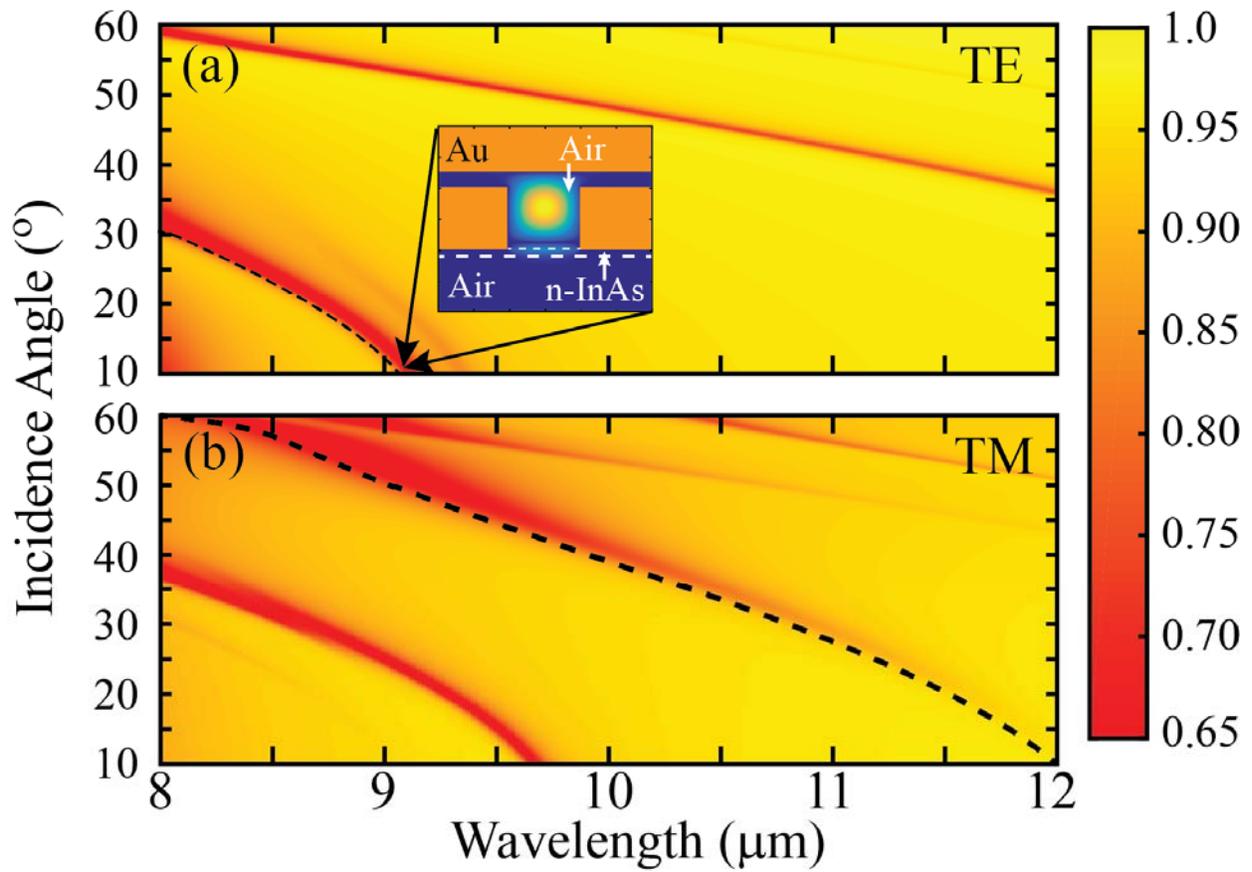

**Figure 5 | RCWA Simulated Coupling to ENZ Photonic Wires.** RCWA-simulated (a) TE and (b) TM polarized reflection as a function of incidence angle. Dashed line in TM-polarized data denotes the calculated coupling angle.

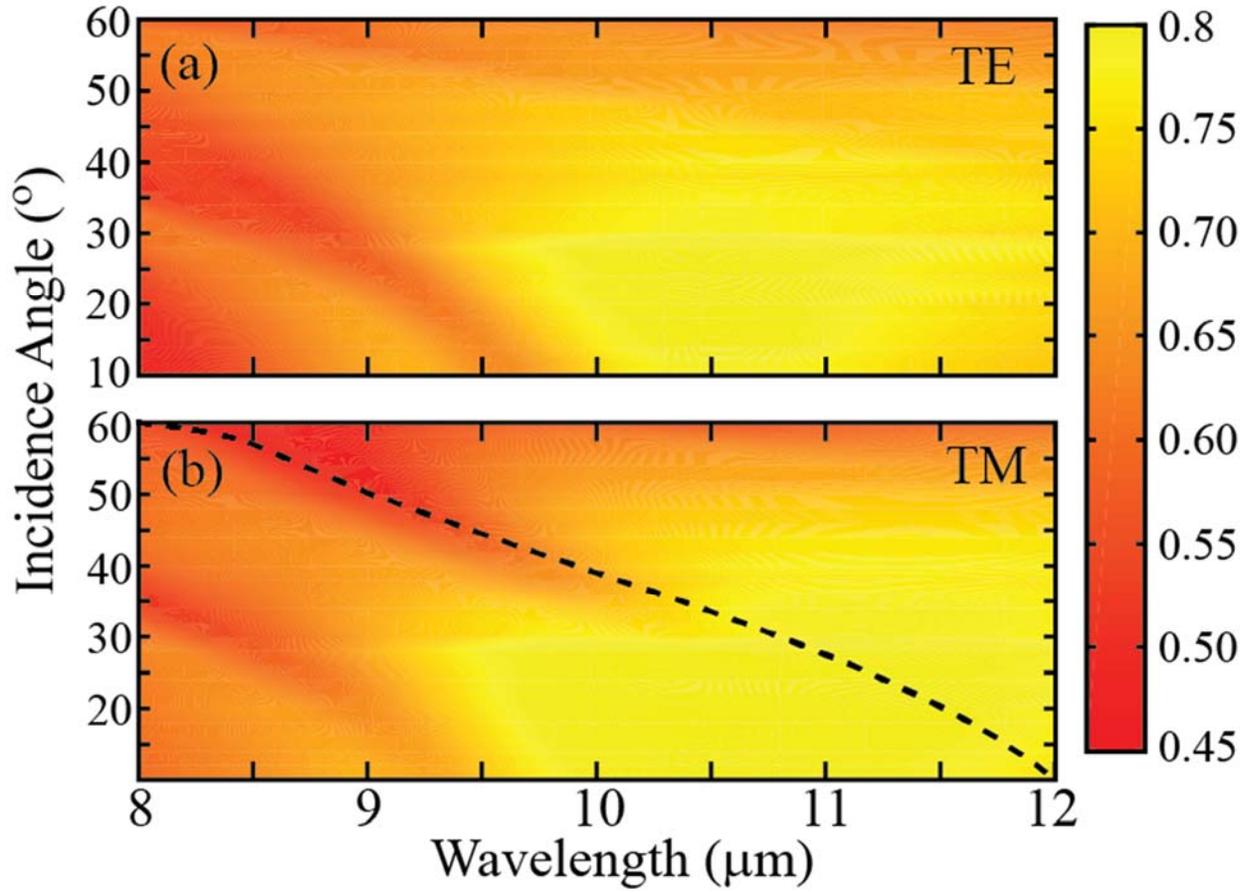

**Figure 6 | Experimental Coupling to ENZ Photonic Wires.** Experimental (a) TE and (b) TM polarized reflection as a function of incidence angle. Dashed line in TM-polarized data denotes the calculated coupling angle.

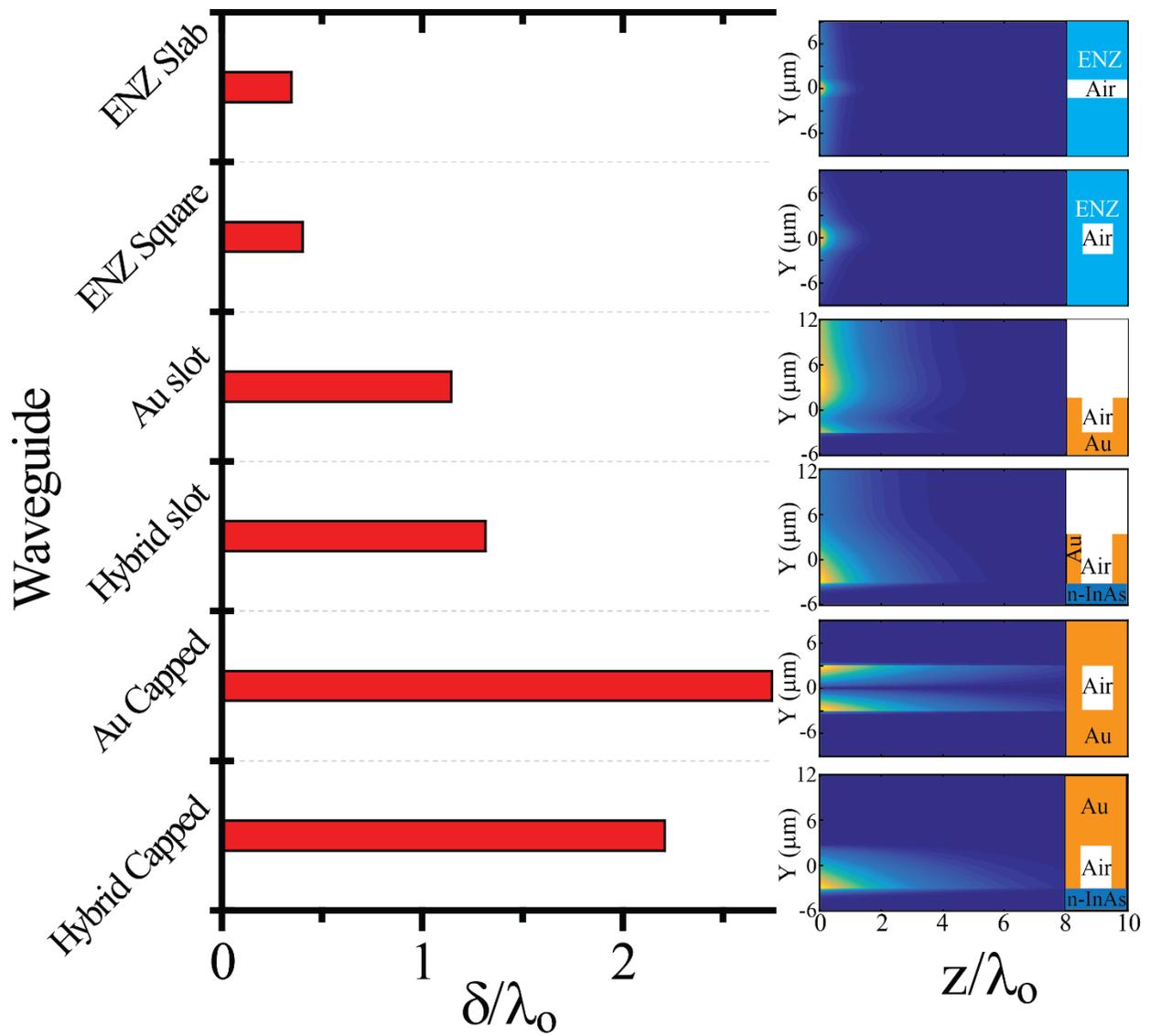

**Figure 7 | Propagation Lengths and Mode Profiles of Photonic Wire Designs**. Comparison of propagation lengths for the potential ENZ waveguides discussed, normalized to free space wavelength. Contour plots show the cross-sectional normalized field profiles ($|E_z|$) for each of the waveguide geometries, along the direction of propagation (normalized to free-space wavelength), with insets showing basic schematics of the waveguide geometry.